# On the Roles of Escape Erosion and the Relaxation of Craters on Pluto

S. Alan Stern, Simon Porter, and Amanda Zangari
Southwest Research Institute
Boulder, CO 80302

## Abstract

Pluto and its satellites will be the most distant objects ever reconnoitered when NASA's New Horizons spacecraft conducts its intensive flyby of this system in 2015. The size-frequency distribution (SFD) of craters on the surfaces in the Pluto system have long been expected to provide a useful measure of the size distribution of Kuiper Belt Objects (KBOs) down to much smaller size scales than presently observed. However, currently predicted escape rates of Pluto's atmosphere suggest that of order one-half to several kilometers of nitrogen ice has been removed from Pluto's surface over geologic time. Because this range of depths is comparable to or greater than most expected crater depths on Pluto, one might expect that many craters on Pluto's surface may have been removed or degraded by this process, biasing the observed crater SFD relative to the production-function crater SFD. Further, if Pluto's surface volatile layer is comparable to or deeper than crater depths, and if the viscosity of this layer surface ice is low like the viscosity of pure $N_2$ ice at Pluto's measured 35 K surface temperature (or as low as the viscosity of $CH_4$ ice at warmer but plausible temperatures on isolated pure-$CH_4$ surfaces on Pluto), then craters on Pluto may also have significantly viscously relaxed, also potentially biasing the observed crater SFD and surface crater retention age. Here we make a first exploration of how these processes can affect the displayed cratering record on Pluto. We find that Pluto's surface may appear to be younger owing to these effects than it actually is. We also find that by comparing Pluto's cratering record to Charon's, it may be possible to estimate the total loss depth of material from Pluto's surface over geologic time, and therefore to estimate Pluto's time-averaged escape rate.

# 1 Introduction

The reconnaissance of the Pluto system by New Horizons in 2015 will shed light on many aspects of this planet and its satellites (e.g., Stern 2008).

Of relevance to our work here, New Horizons imagery of the Pluto system is expected to provide valuable insight into the population distribution of impacting Kuiper Belt Objects (KBOs) via the study of crater size frequency distributions on Pluto and its satellites. Much of this work will be carried out using data from the LOng Range Reconnaissance Imager (LORRI) on New Horizons, which will achieve maximum resolutions of 0.07 km/pixel on Pluto and 0.15 km/pixel at Charon, respectively, with characteristic hemispherical resolutions of 0.46 km and 0.61 km/pixel respectively[1] (Weaver 2014, pers. Comm.).

At such resolutions, craters of diameter greater than about 1 km should be resolved across large expanses of both Pluto and Charon[2]. The parent bodies of such 1 km craters likely correspond to impactor diameters near 100 m (estimated from Eq. 1 below), much smaller than the smallest KBOs currently detectable from Earth. Still smaller craters from still smaller impactors down to several tens meters in diameter may be recognized via their ejecta blankets or in the highest resolution planned New Horizons images. As a result, New Horizons should provide valuable and otherwise unobtainable insights into the KBO size frequency distribution (SFD) at scales from tens of meters up to tens of kilometers in diameter.

---

[1] Owing to the general lack of craters on Triton's surface, it proved to be a poor probe of the Kuiper Belt's population structure (e.g., Stern & McKinnon 2000). Moreover, many craters imaged there were likely from impactors with a planetocentric origin (Schenk & Zahnle 2007).

[2] Imagery of Nix and Hydra will achieve somewhat lower resolutions; this, and their smaller surface areas make them less suitable than Charon for crater size frequency comparison to Pluto, and so these satellites will not be further discussed in this paper.

However, Pluto has an ~10 microbar-class atmosphere (e.g., Elliot et al. 2007). Owing to a combination of Pluto's low gravity and an ~100 K upper atmospheric temperature, Pluto's atmosphere is predicted to be escaping at rates between $10^{27}$ and $10^{28}$ molecules s$^{-1}$ (e.g., Zhu et al. 2014; Tucker et al. 2012; Strobel 2008; Krasnopolsky 1999). Such escape rates, unless only recent or sporadic, imply of order one-half to several kilometers of volatile nitrogen, CO, and CH$_4$ ice have likely been removed from Pluto's surface over geologic time[3]. In what follows we assume that these $10^{27\text{-}28}$ s$^{-1}$ escape rates, quoted widely in the literature, are correct, but note that significantly lower escape rates—even as time averages—would largely negate the effects of escape erosion that we discuss here.

Because this range of depths of surface N$_2$ loss is comparable to or greater than most expected crater depths on Pluto, it is possible that many craters on Pluto's surface may have been largely erased by the loss of surface material to escape. Such an effect would bias the observed crater SFD relative to the native or production-function crater SFD, in turn biasing KBO SFDs. Further, owing to $N_2$'s weak viscosity at the 35 K surface temperature characteristic of Pluto, craters there may also have viscously relaxed, creating geomorphological changes that also affect the observed crater SFD and apparent surface Crater Retention Age (CRE), also biasing the KBO SFD.

---

[3] This in turn implies either internal resupply or a very pure volatile layer. A detailed discussion of these possibilities was first made in Stern et al. (1988).

As a guide to the interpretation of New Horizons imagery, we explore how these two evolutionary processes—crater escape erosion and crater viscous relaxation—may have an affect Pluto's observed cratering record. In what follows, all references to surface age refer to the apparent CRE.

## 2 Methods

In order to simulate the expected crater production population on Pluto and Charon, one needs to define both an expected impactor size distribution and the resulting crater sizes. There is considerable uncertainty in the literature as to the small-end size distribution of Kuiper Belt impactors, so we selected a variety of plausible distributions in the recent literature to represent various possibilities. Like impactor SFDs, crater scaling laws are also not very well defined for low velocity impacts on icy surfaces, so we again selected multiple examples and modeled the resulting crater diameters and depth to diameter ratios. We then modeled the effects of crater erosion and relaxation on our derived impactor SFDs to evaluate the degree to which erosion and relaxation change the observable crater SFDs over time.

**2.1 Impactor size distributions**

Ground-based surveys of KBOs (e.g. Millis et al. 2002; Petit et al. 2011) have shown that KBOs can be grouped into several major dynamical groups. The major such groups are the classical low inclination, low eccentricity KBOs, resonant objects (like Pluto) which have mean motion resonances with Neptune and are more dynamically excited than the classical belt, and the scattered disk of KBOs (like Eris) which are in inclined and eccentric orbits not associated with a mean motion resonance with Neptune. Pluto's orbit is immersed in and has been bombarded by all three populations to varying degrees over time (e.g., Dell'Oro et al. 2013).

The impactor flux on Pluto and Charon will naturally be dominated by smaller KBOs which are too faint to observe from Earth[4]. As a result, the SFD of KBOs smaller than about 10 km can presently only be extrapolated from the SFD of larger KBOs. This is particularly problematic for a range of reasons, including that collisions between KBOs are expected to typically be disruptive, eroding away small KBOs and potentially creating a break in the KB's size frequency distribution (e.g. Stern 1996; Leinhardt et al. 2008, and references therein). Several different predictions have been made of the location of this size break in the KBO population and its effect on the distribution of impactors on Pluto and Charon. Durda & Stern (2000) used the model of Weissman & Levison (1997) with the break at 10 km to derive a total of 8900 impactors larger than 1 km on Pluto over the past 3.5 Gyr. Later, Zahnle et al. (2003) derived a more complex size distribution with multiple breaks at 1.5 km, 5 km, and 30 km, which produced 5250 impactors on Pluto for their Case A and 18,300 impactors larger than 1 km for their Case B. Even more recently, using a numerical collision code with an initial size break at 60 km, de Elía et al. (2010) estimated that over the past 3.5 Gyr Pluto has collided with 1271 to 5552 impactors larger than 1 km. And more recently, Bierhaus & Dones (2014) combined Fraser et al.'s (2014) KBO population model with size a break at 145 km for "cold" low-inclination KBOs and 130 km for "hot" high-inclination KBOs with Pluto collision probabilities from Dell'Oro et al. (2013) to estimate that 350 to 1750 impactors larger than 1 km have hit Pluto over the past 3.5 Gyr. The various estimates in the number of 1 km craters expected on Pluto just reviewed differ by over an order of magnitude, reflecting the significant extant uncertainty in the number of small KBOs.

---
[4] For example, a 10 km KBO even with a high albedo of 0.5 at 40 AU has a visual magnitude of only about 27, while a 1 km KBO at this albedo and distance would have a visual magnitude of 32.

Figure 1 displays predicted Pluto cratering relative size-frequency distributions (called R-plots) for the different Kuiper Belt impact populations discussed just above. The results, shown in Figure 1, assume an impactor density of 500 kg m$^{-3}$, as may be typical of most smaller KBOs (Vilenius et al. 2014, and references therein). With this density, the craters in most cases reach geometric saturation (R=0.2) at around 1 km diameter. If the impactors were all higher density, near 2000-2500 kg m-$^3$ (a plausible near-bounding case in the Kuiper Belt; e.g., Vilenius et al. 2014), then the cratering distributions in Figure 1 would shift upward, but keep the same slopes. In this higher impactor density case, craters would reach saturation at closer to 10 km diameter. In either case there are a sufficient number of smaller primary craters to make identification of secondary craters difficult (Bierhaus & Dones 2014).

Of the nine Kuiper Belt impactor models from the literature plotted in Figure 1, four were adopted to go forward with in this work, so as to reduce unnecessary plot complexity in what follows. These four were chosen on the basis of both their plausibility and representativeness. For example, the three populations of de Elía et al. (2010) ("ESB10") are essentially one main prediction (Population 2), with Populations 1 and 3 representing larger and smaller initial power law indices respectively; hence, their intermediate prediction was chosen. Models "BD14 q=2.00" (Bierhaus & Dones 2014; here and later, q is used to denote the exponent of the population power law size distribution) and "ZSLD03 A" (Zahnle et al. 2003) were chosen because they represent bounding scenarios that do not resemble other predictions in the model set. Although "DS00" Durda & Stern (2000), "ZSLD03 B" (Zahnle et al. 2003) and "BD14 q=2.95" (Bierhaus & Dones 2014) were each based upon different initial KBO populations, they produce quite similar crater populations on Pluto and Charon. We chose the BD14 q=2.95 model as representative, since it better reflects the current knowledge of KBO populations.

## 2.2 Crater scaling from impactor size

Owing to the predominance of impactors from the classical Kuiper Belt, Dell'Oro et al. (2013) estimates a mean impactor approach speed to the Pluto system is near 1.9 km s$^{-1}$. Accounting for gravitational focusing (Krivov et al. 2003), this leads to average impact speeds of 2.3 km s$^{-1}$ and 2.0 km s$^{-1}$ on Pluto and Charon, respectively. By comparison, Zahnle et al. (2003) give average impact speeds of 20 km s$^{-1}$ for Ganymede, 16 km s$^{-1}$ for Rhea, 10.3 km s$^{-1}$ for Ariel, and 8.2 km s$^{-1}$ for Triton. Clearly, impacts on Pluto and Charon occur at much lower velocities than most icy bodies that have been previously explored by spacecraft. The only objects which have been explored by spacecraft and which have comparably low impact velocities to Pluto and Charon are outer irregular satellites of giant planets, such as Phoebe, which has an estimated mean impact speed of 3.2 km s$^{-1}$ (Zahnle et al. 2003).

We estimate crater diameters using the Zahnle et al. (2003) formalism for gravity regime impactors at 45 deg, resulting in the following expressions for icy satellite crater diameter $D$ as a function of impactor diameter $d$, mean impactor density $\rho_i$, and impactor velocity $v$:

$$D_s = 11.9 \, (v^2/g)^{0.217} (\rho_i/\rho_t)^{0.333} d^{0.783} \text{km} \qquad (1)$$

$$D = \begin{cases} D_s : D_s < D_c \\ D_s(D_s/D_c)^\xi : D_s \geq D_c \end{cases} \qquad (2)$$

Here $g$ is the target surface acceleration due to gravity, $\rho_t$ is the mean density of the target surface, $D_c$ is the diameter where the crater transitions from simple to complex, and $\xi$ is a scaling factor that Zahnle et al. (2003) took to be 0.13 after McKinnon et al. (1991). We use Zahnle's relations above for this paper, with surface gravities on Pluto and Charon of 0.66 m s$^{-2}$ and 0.28 m s$^{-2}$, respectively. As discussed above, we assumed a nominal impactor density of 500 kg m$^{-3}$, consistent with measured values for KBOs smaller than 100 km (Grundy et al.

2008), but also examined high-end bounding test cases at 2500 kg m$^{-3}$. To account for a small amount of surface porosity, we assumed a surface bulk density of 900 kg m$^{-3}$ for both Pluto and Charon[5]. Because the typical impact velocity on Pluto is so low, as is the expected bulk density of the impactors, crater diameters on Pluto seem anomalously small for given sized impactors, compared to inner solar system analogs. For example, following Eqn (1) above, a 1 km diameter impactor with a density of 0.5 g cm$^{-3}$ will only produce a transient crater diameter near 6 km for Pluto, versus for example 11 km for Callisto.

Crater depths H are more difficult to estimate. Schenk (1989) showed that they vary widely among the icy satellites, without an easy correlation to surface gravity or impact velocity. We therefore adopted two end-members of the empirical fits to crater depths reported in Schenk (1989). For the deep crater case, we chose a depth scaling similar to the icy Saturnian satellite Rhea:

$$D_c = 13 \text{ km} \qquad (3)$$

$$H = \begin{cases} 0.1\, D : D < D_c \\ 0.366\, D^{0.559} : D \geq D_c \end{cases} \qquad (4)$$

And for the shallow crater case, we chose a depth scaling similar to the icy Uranian satellite Oberon:

---

[5] We note that Bierhaus & Dones (2014) used the model of Housen & Holsapple (2011) to estimate transient crater diameters and volume. However, they assumed the transient crater diameter was equal to the final, and did not include complex craters, resulting in crater diameters that were slightly smaller than Equation 2.

$$D_c = 0.8 \text{ km} \qquad (5)$$

$$H = \begin{cases} 0.1\, D : D < D_c \\ 0.092\, D^{0.671} : D \geq D_c \end{cases} \qquad (6)$$

## 2.3 Modeling crater modification by escape erosion

Pluto has tenuous atmosphere dominated by molecular nitrogen with an estimated surface pressure of order ~0.3 Pa (e.g., Young 2013). Solar ultraviolet heating of Pluto's upper atmosphere drives escape, causing the planet to lose $10^{27}$ to $10^{28}$ $N_2$ s$^{-1}$ (e.g., Zhu et al. 2014, and references therein). This equates to mass loss rates of ~50 to ~500 kg s$^{-1}$. Extrapolating the mass loss rate over 3.5 Gyr and assuming a surface density of 900 kg m$^{-3}$ as noted above, indicates that a global layer of $N_2$ ice some ~0.3 to ~3 km thick has been removed over time at current escape rates[6]. By comparison, current volatile transport models for Pluto predict less than a 1 m thick surface layer of $N_2$ ice involved in annual volatile transport (Young 2013)—negligible compared to escape.

Here we simplistically assume that escape erosion will cause each crater's apparent depth to decrease linearly with time while its shape remains unchanged. This model assumes erosion is happening equally both on the rim and the floor of the crater, and assumes that the depth of the volatile layer on Pluto is comparable to or greater than the crater depth. Since nothing is known about the possibility of layering or the depth of Pluto's surface volatiles at the time of this writing, we believe that more complex models making additional assumptions such as various finite depths of a volatile layer, are not useful at this time.

---

[6] Of note, Triton has an $N_2$ atmosphere with similar surface pressure to Pluto, but an escape rate slower than $10^{24}$ $N_2$ s$^{-1}$ (Lammer 1995). Assuming this rate has been in effect over the age of the Solar System, implies <1 m of nitrogen ice has been lost from its surface over the age of the Solar System. Similarly, We note that the absence of any significant atmosphere on Charon precludes escape erosion from affecting crater SFD on it.

**We model the evolution of the crater depth as:**

$$H(t) = H(0) - 0.1 \left( \frac{F}{\frac{10^{27} N_2}{s}} \right) \left( \frac{t}{Gyr} \right) \text{ km} \qquad (7)$$

where $F$ is the atmospheric escape rate in $N_2$ s$^{-1}$ and $t$ is time in billion years. We apply this to explore the effect of escape erosion on the observed SFD of Pluto's craters.

To do so, we generated suites of synthetic craters for given KB production SFDs shown in Figure 1 with estimated diameters and depth calculated as described above, and then assigned each crater a random age uniformly distributed between 0 and 3.5 Gyr. We then eroded each crater by modifying its depth according to its age, assuming a constant, long-term time-averaged escape rate, and then removed all the craters whose depth of erosion was greater than the initial depth of the crater. For craters that did not completely erode away, we calculated a new diameter and depth/diameter ratio from its post-erosion depth.

This model is oversimplified in that it doesn't recognize that in a volatile layer as deep or deeper than the crater, the bottom of the crater will lose material as will the top level of the planetary surface. However, a model properly accounting for this would require finite element modeling of the differential insolation at the planetary surface and crater bottom, which in turn is dependent on crater depth, diameter, and latitude, which is beyond the scope of this exploratory paper. Therefore we have chosen instead to simply demonstrate the fact that crater escape erosion can make observable effects even when this effect produces erosion only 1-10% as great as the nominal case by the formalism in Eqn (7).

**2.4 Modeling crater viscous relaxation**

If the depth of a weak volatile like $N_2$ is approximately as deep, or deeper than a crater, then as described in Melosh (1989), crater depth $H$ will decrease as a function of time and temperature owing to the viscous relaxation of its terrain under gravitational force as:

$$H(t) = H(0)\exp(-t/t_R) \qquad (8)$$

where the time constant, $t_R$, is given by:

$$t_R \approx \frac{8\eta}{\rho g D} \qquad (9)$$

Here, $\rho$ is the density of the surface, $g$ is the surface gravity, $D$ is the crater diameter, and the effective viscosity is $\eta$. This parameter can be calculated using the Arrhenius relation (e.g., Thomas et al. 1987):

$$\eta_{eff} = \eta_0 \exp\left(24.9\left(\frac{T_m}{T} - 1\right)\right) \qquad (10)$$

where $T_m$ is material's melting point (e.g., 273 K for water ice) and $\eta_0$ is the viscosity at $T_m$, equal to $10^{14}$ Pa-s for water ice. Desch et al. (2009) note that this approximation is valid for grain sizes of less than 1 mm; Buie & Grundy (2000) ruled out grain sizes of anything larger in their Pluto models.

We now employ this formalism to explore crater relaxation on Pluto and Charon, beginning with Charon.

Assuming a $H_2O$-dominated surface and a surface density of 900 kg m-3, 50 K for the surface temperature, and a largest crater size of 300 km, we estimate a $t_R$ of on the order of $10^{55}$ seconds—some $10^{37}$ times the age of the universe. This exceptionally long timescale is due to the water ice dominating Charon's surface being far below its melting point. Smaller craters will relax still more slowly.

However, while viscous crater relaxation on Charon is likely to be negligible currently unless there exist unseen sources of local or global internal heat, Pluto may have experienced much more significant crater relaxation because its surface is dominated by $N_2$ ice (e.g. Owen et al. 1993), which is considerably weaker (less viscous) than $H_2O$ ice. If this $N_2$ is present only as a thin coating on top of underling cold $H_2O$ ice, one would not expect craters on Pluto to be relaxed, for reasons similar to those outlined above for Charon. However, if the $N_2$ ice runs deep into the surface (i.e., comparable to or greater than crater depths), then there could be significant relaxation.

This is because the melting point of $N_2$ ice is 63.15 K, and $\eta_0 \approx 8.8 \times 10^{10}$ Pa-s (Brown & Kirk 1994), giving it a viscosity of approximately $5 \times 10^{15}$ Pa-s at 44 K. Inserting this in Eqn (8) gives a short relaxation timescale of ~2000 years for a 1 km diameter crater and only ~200 years for 10 km crater in deep $N_2$ at Pluto's gravity and surface temperature.

To study the combined effect of escape erosion and crater relaxation, we used a two-step approach. From numerical experiments with Eqns (8)-(10), we find that most crater relaxation will occur either quickly after the crater is formed (i.e., within a few million years) or not at all. In contrast, significant escape erosion only occurs over much longer timescales of several hundred million to several billion years for crater sizes and depth detectable by New Horizons. We therefore took the same synthetic impact crater sets described above, modified their depths appropriately for relaxation, and then applied escape erosion as above, preserving the post-relaxation depth/diameter. As stated above, we assume craters are erased when eroded to their depth or relaxed by on their Maxwell timescale. That idealized "complete erasure" will not actually occur, as geomorphological evidence of relaxed and eroded craters will manifest themselves instead as severely degraded.

## 3 Results

### 3.1 Pluto escape erosion acting alone

Figure 2 shows the results of our calculations after 3.5 Gyr of escape erosion at atmospheric escape rates of $1 \times 10^{27}$ sec$^{-1}$ ("minimum" case), $3 \times 10^{27}$ sec$^{-1}$ ("expected" case), and $1 \times 10^{28}$ sec$^{-1}$ ("maximum" case), as well as comparison curves at 0% (no escape, production function), and 1% and 10% of the $3 \times 10^{27}$ ("mid-range") case to test the likely true range of differential escape erosion effects that modify the crater relative to the planetary surface. We depict all the results as R-plots.

Computations are shown for both shallow depth-to-diameter ratios taken from the Saturnian satellite Rhea (Schenk 1989), and deeper depth-to-diameter ratios taken from the Uranian satellite Oberon (Schenk 1989).

From Figure 2, which is computed for escape erosion computed without any viscous relaxation, we see that escape erosion significantly modifies the R-plots, independent of production function and initial depth-to-diameter case. However, if differential erosion is only taking place at the 1% to 10% of the rates of surface-only erosion discussed above, then the R-plots will not be significantly modified. What is actually happening on Pluto may involve one extreme or the other, so computations like those presented in Figure 2 can serve as a guide to interpretation of R-plots once New Horizons imagery is available.

**3.2 Pluto crater evolution with both escape erosion and relaxation**

We now examine the combined effects of escape erosion and viscous relaxation on the crater SFD on Pluto. We use the crater escape erosion and crater relaxation models described in §2.3 and §2.4 above, respectively. Relaxation plus escape erosion results are shown in Figure 3 (for near-surface T=35 K), Figure 4 (for T=40 K), and Figure 5 (for T=45 K).

Examining these figures we first conclude that even at T=35 K, the expected temperature of the $N_2$-rich regions, large craters in deep $N_2$ will be relaxed sufficiently to be severely or even completely depleted, almost independent of which KBO population and crater depth-to-diameter cases are assumed. We also find that escape erosion dominates the modification at smaller craters sizes, and a combination of escape erosion and viscous relaxation affect the larger craters. For surfaces at T=40 K and T=45 K, we find that relaxation effects are more dramatic, depressing the entire crater size frequency distribution to the point of removing virtually all but the youngest small (and in some cases intermediate) sized craters, regardless of whether the initial depth-to-diameter regime is deep or shallow and independent of KBO population.

Figure 6 extends these results, depicting crater relaxation timescales as a function of crater size on Pluto for deep $N_2$, and also for deep $CH_4$ surfaces, which may be relevant at some locations on Pluto. The viscosity of $CH_4$ is intermediate to $H_2O$ and $N_2$; we adopted a value of $2 \times 10^9$ Pa-s at $T_m$=89.5 K (Yamashita and Kato 1997) [7]. Although $CH_4$ does not produce significant relaxation at T=35 and T=40 K, one can see that $CH_4$ surfaces could result in significant relaxation for temperatures of 45 K or higher.

As a consequence of the results presented here, it is possible, particularly if (or where) the warmer temperature cases apply, that Pluto's crater SFD may mimic a young surface even if the surface is old.

One telltale signature that can be employed to aid in breaking the ambiguity between a relaxation-evolved surface and a truly young (i.e., recently created) surface would be the presence or absence of tectonics, flow fronts, and other evidence of internal activity in units with few craters. Another would be the detection in New Horizons imagery of craters that show geomorphological signs of relaxation. Yet another would be that relaxation processes will deform and degrade both negative and positive topographic features, i.e., both craters and mountains and scarps.

---

[7] We were not able to locate viscosities for CO ice at the relevant temperatures.

As an important aside, if as one expect we can recover a native (i.e., un-eroded) R-plot from New Horizons imagery of Charon, then by matching observed Pluto SFD to model runs like these, using as inputs Charon R-plots adjusted for g and impact velocity to derive the production crater SFD on Pluto, it may be possible from the small-medium scale crater deficit to derive an integral estimate of the long-term integral atmospheric escape rate of Pluto over billions of years[8].

## 4 Summary

We have examined how the size frequency distribution (SFD) of craters detected on Pluto can significantly evolve owing to volatile escape erosion acting alone or in concert with surface-temperature-dependent viscous relaxation.

---

[8] However we note that if the volatile layer on Pluto is not deep compared to the crater depths, then derivation of a time integral escape rate will be significantly more problematic.

We found that if the volatile layer seen on Pluto's surface is deep, then over a wide range of possible impactor SFDs, crater depth-to-diameter ratios, and surface temperatures, the evolutionary effects of escape erosion and viscous relaxation can bias the estimation of both the Kuiper Belt impactor size frequency distribution and crater retention ages. This however will likely not be the case if the volatile is only a veneer above an $H_2O$-dominated topography[9]. Further, we caution that past claims (e.g., Bierhaus & Dones 2014) that Pluto's surface will reflect the KB SFD itself should be considered necessarily valid only if the volatile layer is shallow.

Finally, we discussed the reasons that that neither escape erosion nor relaxation are likely to be important on Charon. We therefore described how it may be possible to use the comparative cratering records of Pluto and Charon to determine the integral loss depth of material from Pluto's surface, and therefore the integral escape loss from the atmosphere, over geologic time.

**Acknowledgements**

This work was supported by NASA's New Horizons mission under contact NASW-02008 to the Southwest Research Institute. We thank Hal Weaver, Clark

---

[9] Our crater relaxation conclusions may also have relevance to crater evolution on Triton if the volatile layer there is deep.

Chapman and Bill McKinnon for useful conversations; we also thank two anonymous referees and Beau Bierhaus, Luke Dones, Dan Durda, Jeff Moore, Isaac Smith, and Kelsi Singer for constructive critiques of this manuscript.

# References

Bierhaus, E.B., and Dones, H., 2014. Craters and ejecta on Pluto and Charon: Anticipated results from the New Horizons flyby. Icarus (in press).

Brown, R.H., and Kirk, R.L., 1994. Coupling of volatile transport and internal heat flow on Triton. J. Geophys. Res., 99, 1965–1981. doi:10.1029/93JE02618.

Buie, M.W., and Grundy, W.M., 2000. The distribution and physical state of $H_2O$ on Charon. Icarus, 148, 324–339. doi:10.1006/icar.2000.6509.

de Elía, G.C., di Sisto, and R.P., Brunini, A., 2010. Impactor flux and cratering on the Pluto-Charon system. Astron. & Astrophys., 521, A23-27. doi:10.1051/0004-6361/201014884, arXiv:1007.0415.

Dell'Oro, A., Campo Bagatin, A., Benavidez, P.G., and Alemañ, R.A., 2013. Statistics of encounters in the Trans-Neptunian region. Astron. & Astrophys. 558-565, A95. doi:10.1051/0004-6361/201321461.

Desch, S.J., Cook, J.C., Doggett, T.C., and Porter, S.B., 2009. Thermal evolution of Kuiper Belt Objects, with implications for cryovolcanism. Icarus, 202, 694–714. doi:10.1016/j.icarus.2009.03.009.

Durda, D.D., and Stern, S.A., 2000. Collision rates in the present-day Kuiper Belt and centaur regions: Applications to surface activation and modification on comets, Kuiper Belt Objects, Centaurs, and Pluto-Charon. Icarus, 145, 220–229. doi:10.1006/icar.1999.6333, arXiv:astro-ph/9912400.

Elliot, J.L., Person, M.J., Gulbis, A.A.S., Souza, S.P., Adams, E.R., Babcock, B.A., Gangestad, J.W., Jaskot, A.E., Kramer, E.A., Pasachoff, J.M., Pike, R.E., Zuluaga, C.A., Bosh, A.S., Dieters, S.W., Francis, P.J., Giles, A.B., Greenhill, J.G., Lade, B., Lucas, R., and Ramm, D.J., 2007. Changes in Pluto's atmosphere: 1988-2006. Astron. J., 134, 1–13. doi:10.1086/517998.

Fraser, W.C., Brown, M.E., Morbidelli, A., Parker, A., and Batygin, K., 2014. The absolute magnitude distribution of Kuiper Belt Objects. Astrophys. J., 782, 100-114. doi:10.1088/0004-637X/782/2/100, arXiv:1401.2157.

Grundy, W.M., Noll, K.S., Virtanen, J., Muinonen, K., Kern, S.D., Stephens, D.C., Stansberry, J.A., Levison, H.F., and Spencer, J.R., 2008. (42355) Typhon Echidna: Scheduling observations for binary orbit determination. Icarus, 197, 260–268. doi:10.1016/j.icarus.2008.04.004, arXiv:0804.2495.

Housen, K.R., and Holsapple, K.A., 2011. Ejecta from impact craters. Icarus, 211, 856–875. doi:10.1016/j.icarus.2010.09.017.

Krasnopolsky, V.A., 1999. Hydrodynamic Flow of $N_2$ from Pluto, JGR, 104, 5955-5962.

Krivov, A.V., Sremčević, M., Spahn, F., Dikarev, V.V., and Kholshevnikov, K.V., 2003. Impact-generated dust clouds around planetary satellites: Spherically symmetric case. Planet. Space Sci., 51, 251–269. doi:10.1016/S0032-0633(02)00147-2.

Lammer, H., 1995. Mass loss of N molecules from Triton by magnetospheric plasma interaction. Planet. Space Sci., 43, 845–850. doi:10.1016/0032-0633(94)00214-C.

Leinhardt, Z.M., Stewart, S.T., and Schultz, P.H., 2008. Physical effects of collisions in the Kuiper Belt. In *The Solar System Beyond Neptune,* (M.A. Barrucci, H. Boenhardt, D.P. Cruikshank, and A. Morbidelli, eds.), U. Az. Press, pp. 195–211.

McKinnon, W.B., Chapman, C.R., Housen, K.R., 1991. Cratering of the Uranian satellites. In *Uranus* (J. Bergstralh, and E.D. Miner, eds.) University Az. Press, pp. 629-692.

Melosh, H.J., 1989. *Impact cratering: A geologic process*. Oxford University Press, 245 pp.

Millis, R.L., Buie, M.W., Wasserman, L.H., Elliot, J.L., Kern, S.D., and Wagner, R.M., 2002. The Deep Ecliptic Survey: A search for Kuiper Belt Objects and Centaurs. I. Description of methods and initial results. Astron. J., 123, 2083–2109. doi:10.1086/339481.

Owen, T.C., Roush, T.L., Cruikshank, D.P., Elliot, J.L., Young, L.A., de Bergh, C., Schmitt, B., Geballe, T.R., Brown, R.H., and Bartholomew, M.J., 1993. Surface ices

and the atmospheric composition of Pluto. Science, 261, 745–748. doi:10.1126/science.261.5122.745.

Petit, J.M., Kavelaars, J.J., Gladman, B.J., Jones, R.L., Parker, J.W., Van Laerhoven, C., Nicholson, P., Mars, G., Rousselot, P., Mousis, O., Marsden, B., Bieryla, A., Taylor, M., Ashby, M.L.N., Benavidez, P., Campo Bagatin, A., and Bernabeu, G., 2011. The Canada-France Ecliptic Plane Survey—Full data release: The orbital structure of the Kuiper Belt. Astron. J., 142, 131-190. doi:10.1088/0004-6256/142/4/131, arXiv:1108.4836.

Schenk, P.M., 1989. Crater formation and modification on the icy satellites of Uranus and Saturn—Depth/diameter and central peak occurrence. J. Geophys. Res. , 94, 3813–3832. doi:10.1029/JB094iB04p03813.

Schenk, P.M., and Zahnle, K., 2007. On the negligible surface age of Triton. Icarus, 192, 135–149. doi:10.1016/j.icarus.2007.07.004.

Stern, S.A.; Trafton, L.M., and Gladstone, G.R., 1988. Why is Pluto bright? Implications of the albedo and lightcurve behavior of Pluto. Icarus, 75, 1988, 485-498.

Stern, S.A, 1996. Collisional timescales and the architecture of the ancient, massive Kuiper Disk. S.A. Stern. AJ, 112, 1203-1211.

Stern, S.A., 2008. New Horizons: NASA's Kuiper Belt mission. In *The Solar System Beyond Neptune,* (M.A. Barrucci, H. Boenhardt, D.P. Cruikshank, and A. Morbidelli, eds.), U. Az. Press, p557-562.

Stern, S.A., and McKinnon, W.B., 2000. Triton's surface age and impactor flux revisited. AJ, 119, 945-952.

Strobel, D. F., 2008. $N_2$ Escape rates from Pluto's atmosphere. Icarus, 193, 612-619.

Thomas, P.J., Reynolds, R.T., Squyres, S.W., and Cassen, P.M., 1987. The viscosity of Miranda, In: Lunar and Planetary Institute Science Conference Abstracts, pp. 1016-1017.

Tucker. O.J., J.T. Erwin, J.I. Deighan, A.N. Volkov, and R.E. Johnson, 2012. Thermally Driven Escape from Pluto's atmosphere: A combined fluid/kinetic model. Icarus, 217, 408-415.

Vilenius, E., C. Kiss, T. Müller, M. Mommert, P. Santos-Sanz, A. Pál, J. Stansberry, M. Mueller, N. Peixinho, E. Lellouch, S. Fornasier, A. Delsanti, A. Thirouin, J. L. Ortiz, R. Duffard, D. Perna, and F. Henry, 2014. TNOs are Cool: A survey of the Trans-Neptunian region—X. Analysis of Classical Kuiper Belt Objects from Herschel and Spitzer Observations. Astron. and Astrophys, 564, A35-52. DOI: http://dx.doi.org/10.1051/0004-6361/201322416

Weissman, P.R., and Levison, H.F., 1997. The population of the Trans-Neptunian region: The Pluto-Charon environment. In *Pluto and Charon* (S.A. Stern and D.J. Tholen, eds.), U. Az. Press, pp 559-604.

Yamashita. Y., and Kato, M., 1997. Viscoelastic properties of polycrystalline solid methane and carbon dioxide. GRL, 24, 1227-1330.

Young, L.A., 2013. Pluto's seasons: New predictions for New Horizons. Astrophys. J. Let., 766, L22. doi:10.1088/2041-8205/766/2/L22, arXiv:1210.7778.

Zahnle, K., Schenk, P., Levison, H., and Dones, L., 2003. Cratering rates in the outer Solar System. Icarus, 163, 263–289. doi:10.1016/S0019-1035(03)00048-4.

Zhu, X., Strobel, D.F., and Erwin, J.T., 2014. The density and thermal structure of Pluto's atmosphere and associated escape processes and rates. Icarus, 228, 301–314. doi:10.1016/j.icarus.2013.10.011.

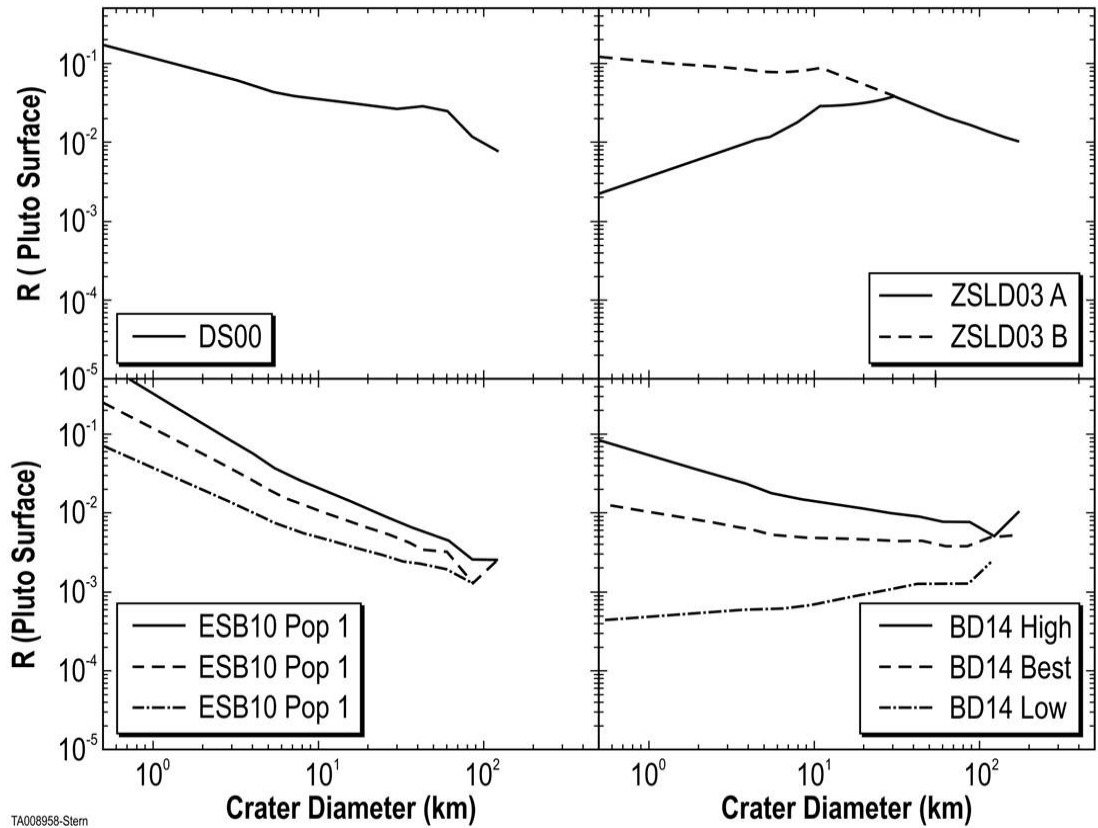

Figure 1: Predicted Pluto crater relative size-frequency model distributions for the different KB impactor populations at a surface age of 3.5 Gyr. "DS00" is from Durda & Stern (2000); "ZSLD03" are from Zahnle et al. (2003); "ESB10" are from de Elía et al. (2010); "BD14" are from Bierhaus & Dones (2014). A surface density of 900 kg m$^{-3}$, an impactor density of 500 kg m$^{-3}$, impactor speeds of 2.3 km s$^{-1}$, and a simple to complex transition of 0.8 km were assumed. Saturation effects are not been included; but saturation occurs at a constant R=0.2. Subsequent figures use models ZSLD03 A, ESB10 Pop 2, BD14 Low, and BD14 High.

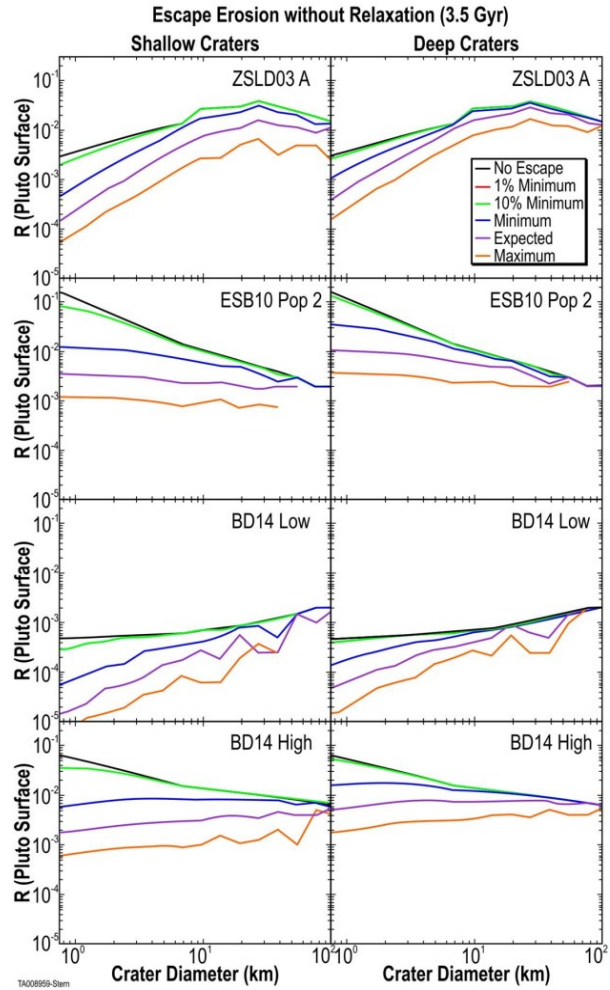

Figure 2: R-plot of 3.5 Gyr of synthetic Pluto crater production SFD (black line), with modified SFDs (colored lines) due to escape erosion only after 3.5 Gyr of escape erosion at atmospheric escape rates of $1\times10^{27}$ sec$^{-1}$ ("minimum" case run), $3\times10^{27}$ sec$^{-1}$ ("expected"), and $1\times10^{28}$ sec$^{-1}$ ("maximum" case run), as well as comparison curves at 0% (no escape, production function), and 1% and 10% of the $3\times10^{27}$ ("mid-range") case to test the likely true range of differential escape erosion effects that modify the crater relative to the planetary surface.

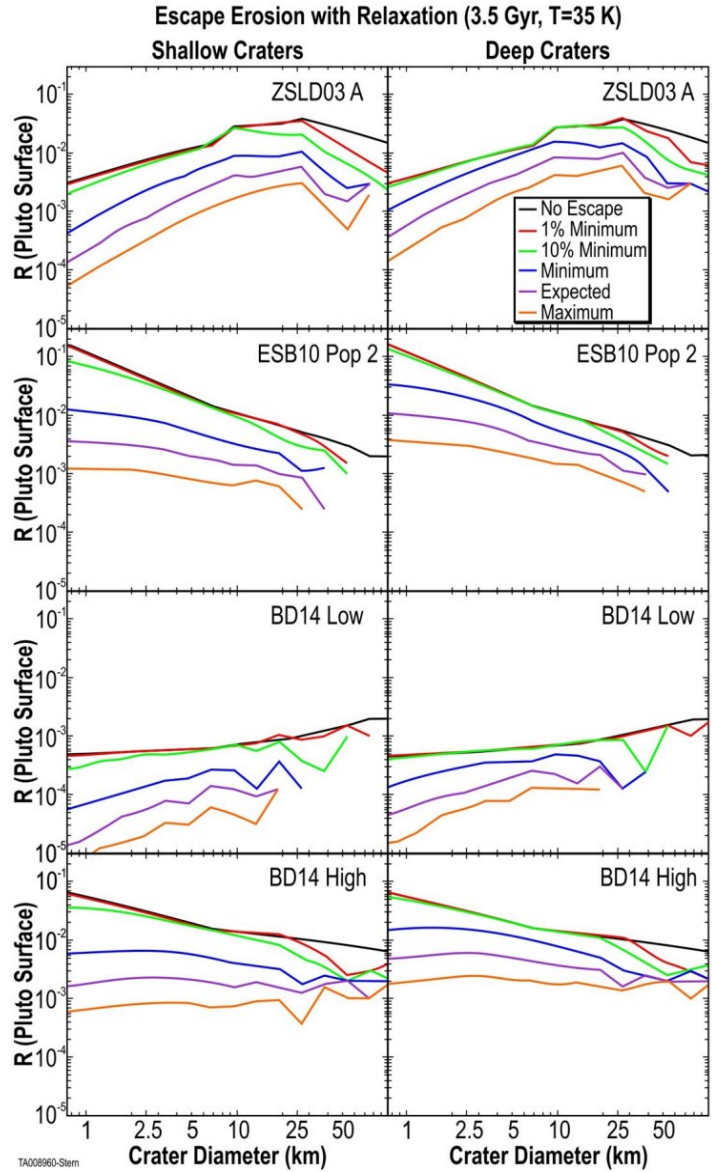

Figure 3: R-plot of 3.5 Gyr of synthetic Pluto crater production SFD (black line), with modified SFDs (colored lines) due to relaxation and escape erosion at T=35 K.

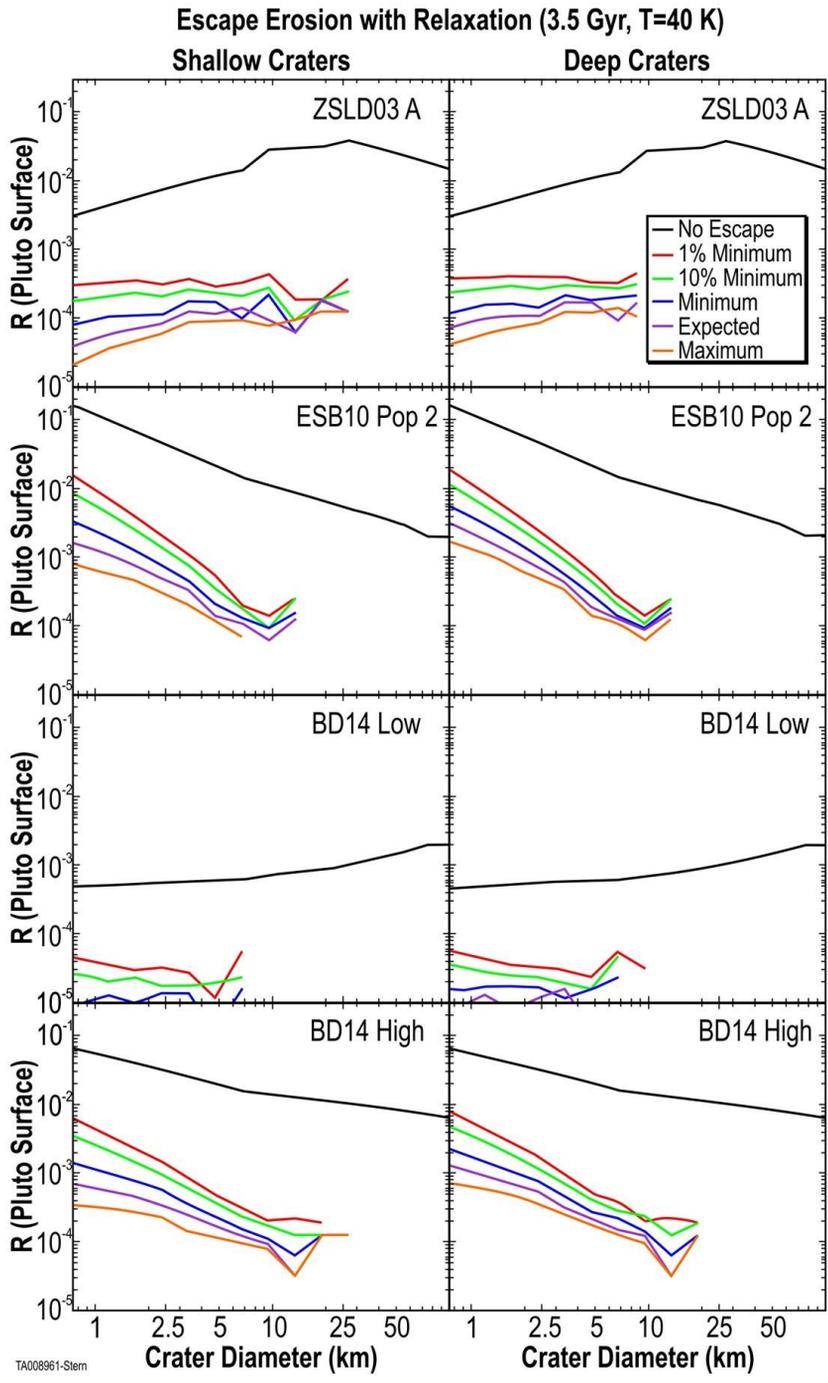

Figure 4: R-plot of 3.5 Gyr of synthetic Pluto crater production SFD (black line), with modified SFDs (colored lines) due to relaxation and escape erosion at T=40 K.

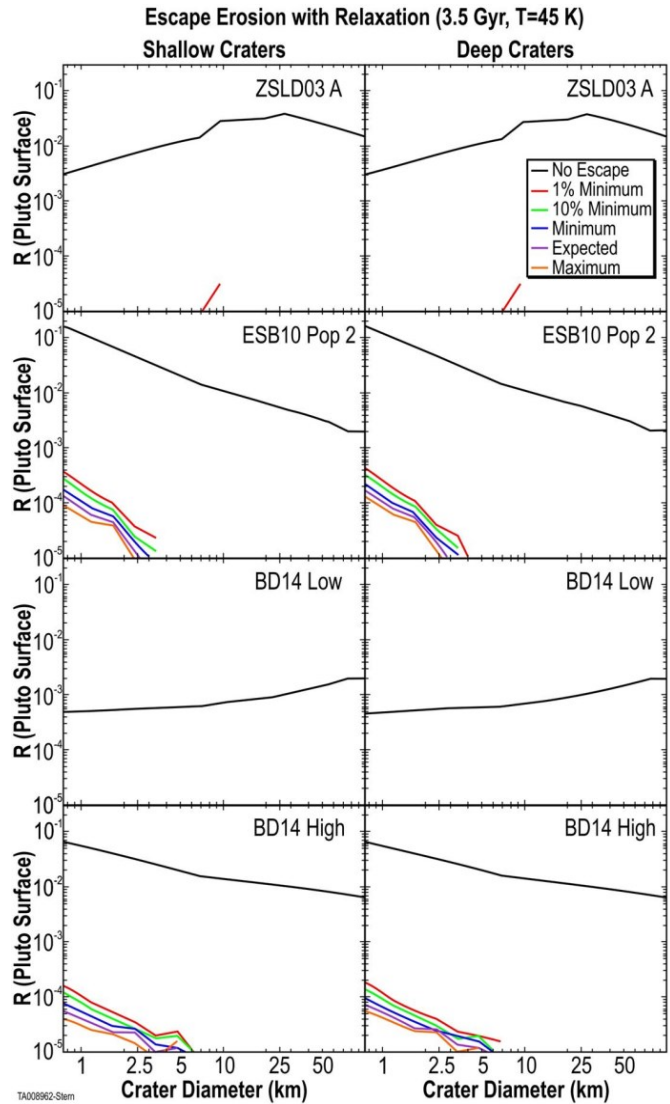

Figure 5: R-plot of 3.5 Gyr of synthetic Pluto crater production SFD (black line), with modified SFDs (colored line) due to relaxation and escape erosion at T=45 K. For the ZSLD03 A and BD14 Low cases, the absence of a colored line indicates a population smaller than the limits of the chart. One should consider the surfaces in these scenarios as predicted to be essentially craterless.

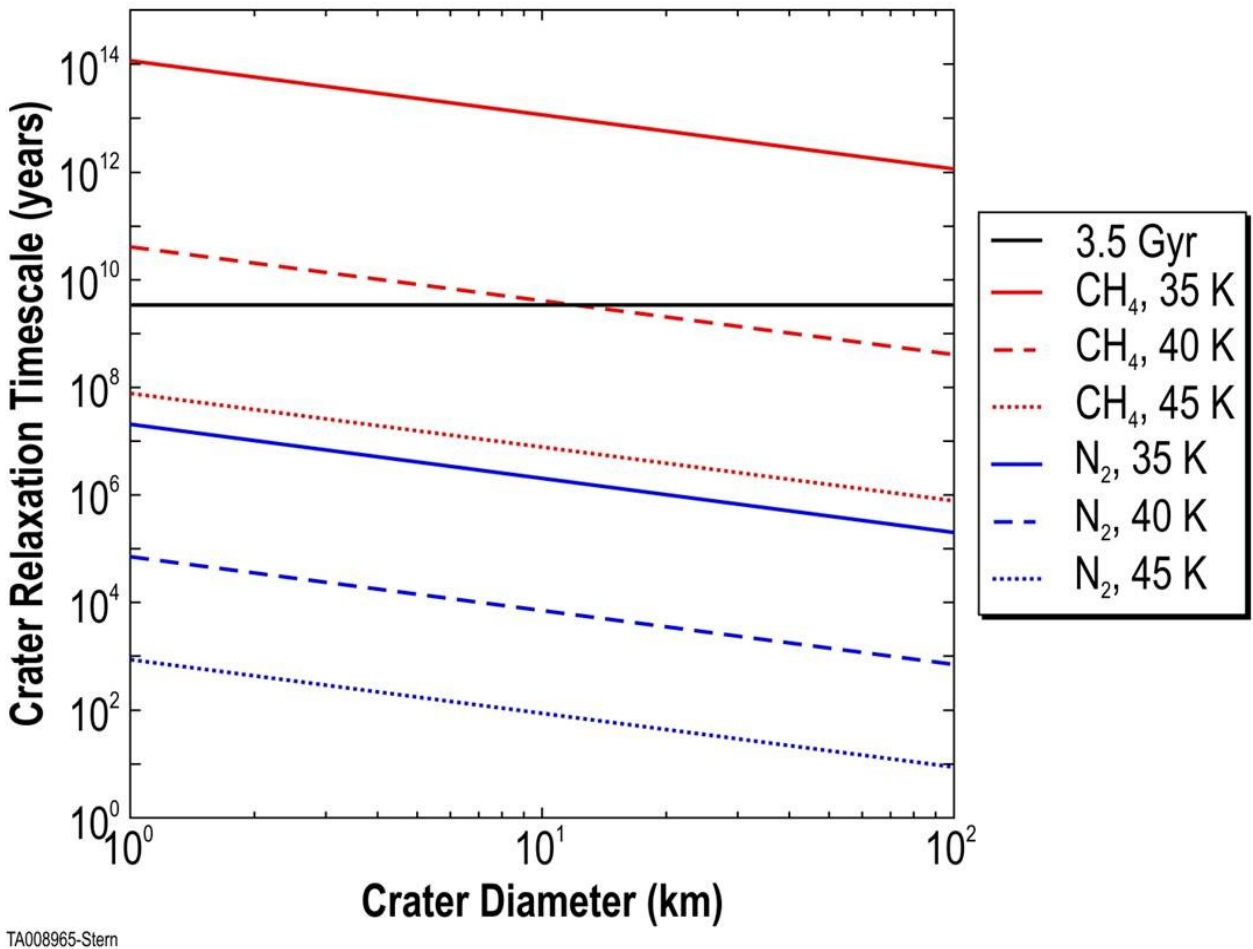

Figure 6: Crater relaxation timescales in $N_2$ and $CH_4$ as a function of temperature. The horizontal black line is at a timescale of 3.5 Gyr.